\documentclass{PoS}

\abstract{Since the detection of high energy astrophysical neutrinos in IceCube, there has been a search for their sources. Although recent evidence of neutrinos from a flaring blazar could explain some of the
neutrino flux, sources for the remainder are still unknown. This analysis searches for neutrinos produced via interactions between diffuse intergalactic Ultra-High Energy Cosmic Rays (UHECR) and matter. In this work the local galaxy density serves as the target for cosmic ray interactions, thus neutrinos produced from these interactions are expected to trace the galaxies spatially. The spatial distribution of galaxies within the local universe (z < 0.10) as seen in the 2MASS Redshift Survey (2MRS) is anisotropic. Here we present an analysis that searches for the spatial correlation between the arrival directions of neutrinos observed at the IceCube neutrino observatory and the directions of high galaxy density in the local universe. No such correlation was found and this analysis presents limits on the flux of neutrinos from the local universe.\\

\vspace{4mm}
{\bfseries Corresponding authors:}
\speaker{Stephen Sclafani}$^{1}$, Naoko Kurahashi Neilson$^{1}$\\
{$^{1}$ \itshape Drexel University}\\

}

\FullConference{36th International Cosmic Ray Conference -ICRC2019-\\
		July 24th - August 1st, 2019\\
		Madison, WI, U.S.A.}

\title{Correlation of IceCube neutrinos with the 2MASS Redshift Survey}

\ShortTitle{Correlation of IceCube neutrinos with 2MASS Redshift Survey}

\author{
The IceCube Collaboration\footnote{For collaboration list, see PoS(ICRC2019) 1177.}\\
{\itshape \href{http://icecube.wisc.edu/collaboration/authors/icrc19_icecube}{http://icecube.wisc.edu/collaboration/authors/icrc19\_icecube}}\\
E-mail: \email{ssclafani@icecube.wisc.edu}
}

\begin{document}

\section{Introduction}

IceCube, a cubic kilometer detector at the geographic south pole, has discovered a flux of astrophysical neutrinos \cite{Aartsen:2014gkd,Aartsen:2016xlq}, and more recently presented compelling evidence for one neutrino source \cite{IceCube:2018dnn, IceCube:2018historicsearch}. However the sources for the majority of the astrophysical neutrino flux remain undefined.  Existing IceCube searches have focused on intrinsic neutrino emission of bright point-like sources, such as blazars \cite{IceCube:2016oji}, or diffuse emission of neutrinos from our own galaxy \cite{IceCube:galactic}. This work, instead, searches for neutrinos produced in interactions between the matter of nearby galaxies and an isotropic intergalactic diffuse flux of UHECR. If such interactions were responsible for significant neutrino production, the spatial distribution of arrival directions of neutrinos would follow the local matter distribution and therefore directional correlation would be observable between the overdensities of the local universe and excesses of astrophysical neutrinos.

\section{2MRS}
\begin{figure}[ht] 
\begin{tabular}{cc}
\includegraphics[width=.48\textwidth]{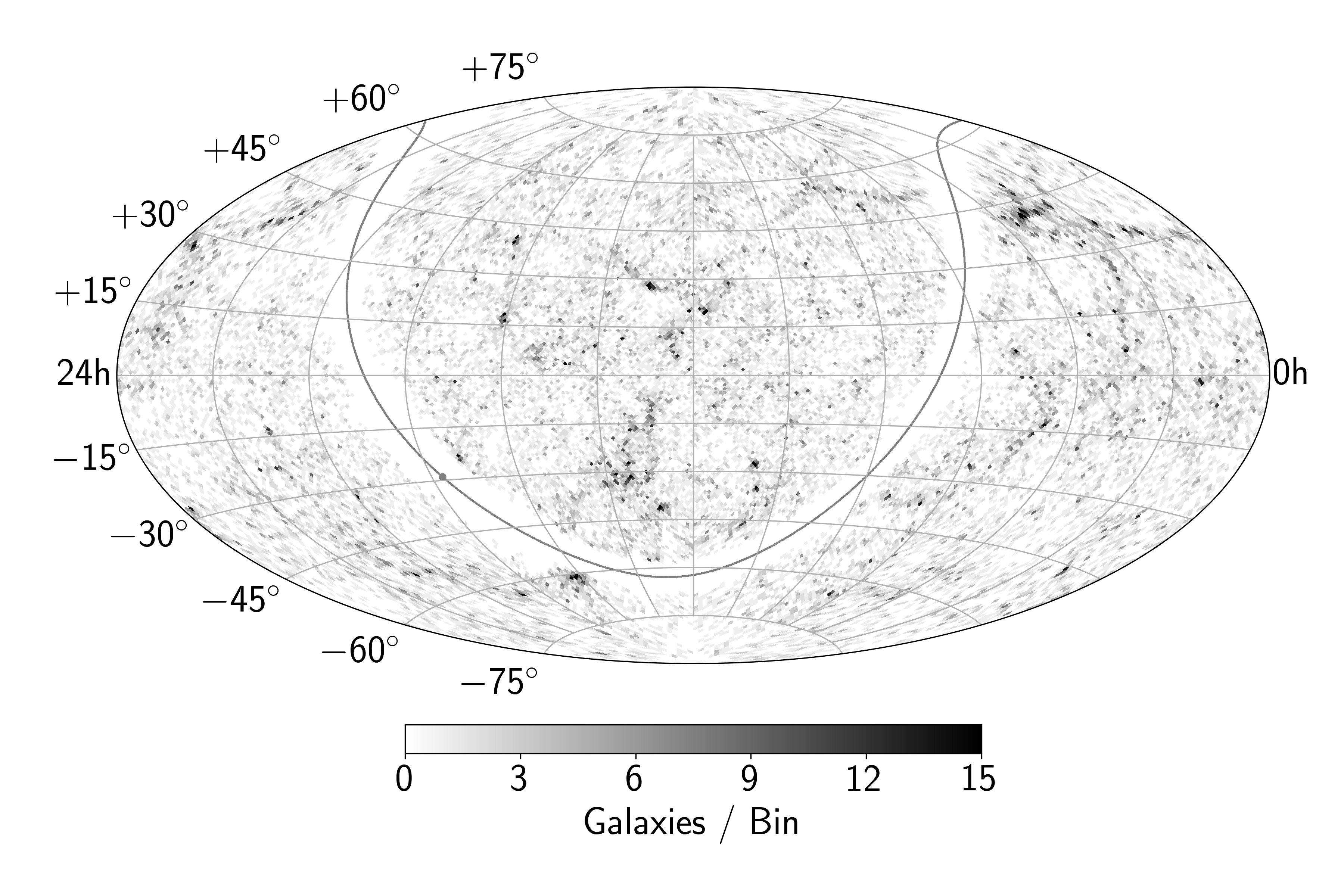}  
& \includegraphics[width=.48\textwidth]{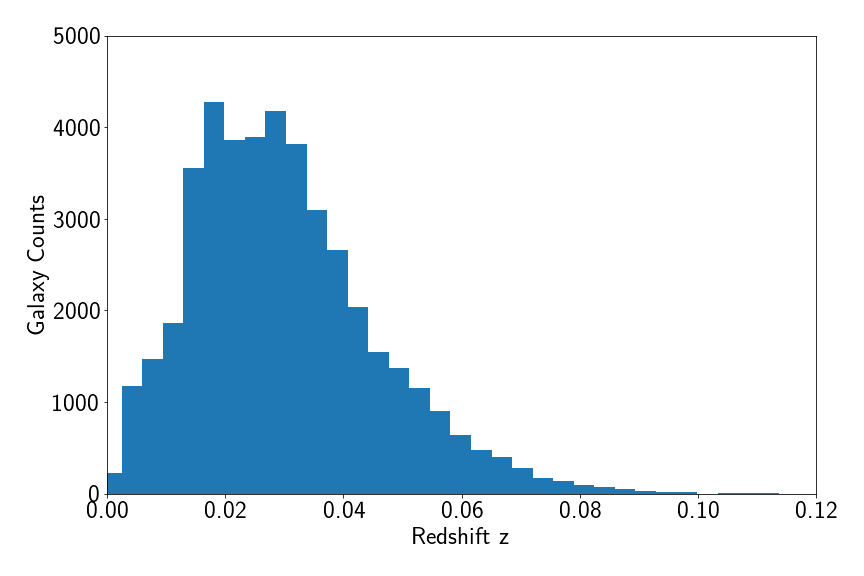} \\
\end{tabular}
\caption{Left: Galaxy density of the complete 2MRS Survey. Right: Histogram of galaxies measured redshift, showing the completeness and ultimate magnitude limit as a function of redshift}
\label{fig:2mrs_density_hist}
\end{figure}

The 2MASS redshift survey (2MRS) is an all-sky catalog of local galaxies with associated redshift measurements, and is the most extensive and unbiased survey up to redshift of $z <  0.03$ \cite{2mrs}.  The infrared survey contains the position, redshift, and $K_s$ band magnitude of 44,000 of the nearest galaxies, including galaxies out to a redshift of $z \approx 0.10$.  The survey is mostly complete to a redshift of $z=0.03$, after which it becomes magnitude-limited. The catalog density as a function of direction and redshift is shown in Figure \ref{fig:2mrs_density_hist}. This analysis uses the 2MRS galaxy catalog to trace the distribution of matter in the local universe.

\section{Analysis}
A neutrino dataset consisting of 7 years of IceCube muon neutrino events previously used in multiple neutrino source searches \cite{IceCube:7yrPS} is also used for this analysis.  A Test Statistic (TS) is defined, describing the similarity between the spatial distribution of neutrinos to the spatial template of galaxy densities obtained from 2MRS. The energies of the neutrino events are also considered in the TS. Since atmospheric neutrinos follow an energy spectrum of $E^{-3.7}$ and astrophysical neutrinos are assumed to follow $E^{-2}$, neutrinos with higher energies are more likely to be astrophysical in origin and assigned higher weight. The background events in IceCube, atmospheric neutrinos and muons, are isotropically distributed within a declination band due to the detector's location at the geographic South Pole. Thus the statistical significance of the correlation found in the data is quantified by comparing the TS of the true data to a distribution of TS values obtained from many background-only data sets where events are uniformly distributed in right ascension.\\

The analysis TS defined in \ref{eq:2} is a likelihood ratio made up of likelihood $L$ \cite{Braun:2008}, which is a function of $n_s$ the number of signal neutrinos. $n_s$ is free to vary between 0 and the total number of observed events, N. The entire dataset is fit for the most likely $n_s$.  
\begin{equation} \label{eq:2}
\mathrm{TS} = -2\ln{\left(\frac{L(n_s=0)}{L(\hat{n_s})}\right)} \\
\end{equation}
where $L(n_s=0)$ is the likelihood for the hypothesis corresponding to no correlation, and $L(\hat{n_s})$ is the likelihood for the best fit $n_s$. \\
 \begin{equation} \label{eq:1}
L(n_s) = \prod_{i=1}^{N}\left( \frac{n_s}{N} S_i(\mathbf{x}_i, \sigma_i, E_i) + (1-\frac{n_s}{N})B_i(\mathbf{x}_i, E_i) \right)
\end{equation} 
The likelihood is defined in \ref{eq:1}. $S_i$ is the likelihood of event $i$ being from the source, and $B_i$ is the likelihood of event $i$ being from the background, which are functions of event direction $\mathbf{x}_i$, energy $E_i$ and angular uncertainty $\sigma_i$.  The signal and background probability density functions (PDF), $S_i$ and $B_i$, follow established  methods  applied previously in IceCube extended template analyses, with $S_i$ calculated using a fixed spectral index of $E^{-2}$.  Detailed description of the method are described in \cite{IceCube:galactic}. \\

\section{Signal Hypothesis}\label{template:PDF}

\begin{figure}[htb]
\begin{tabular}{cc}
\includegraphics[width=.5\textwidth]{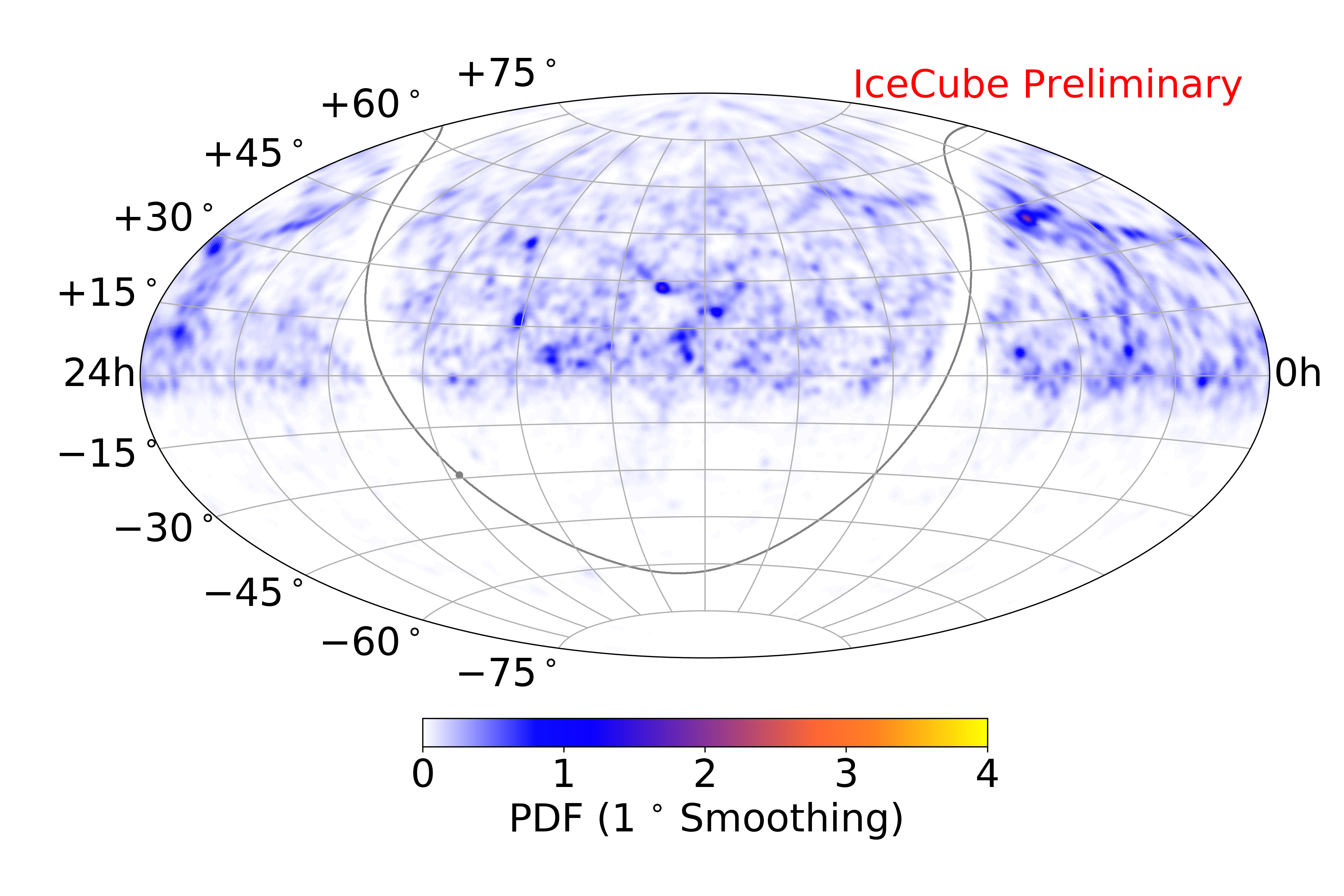} & 
\includegraphics[width=.5\textwidth]{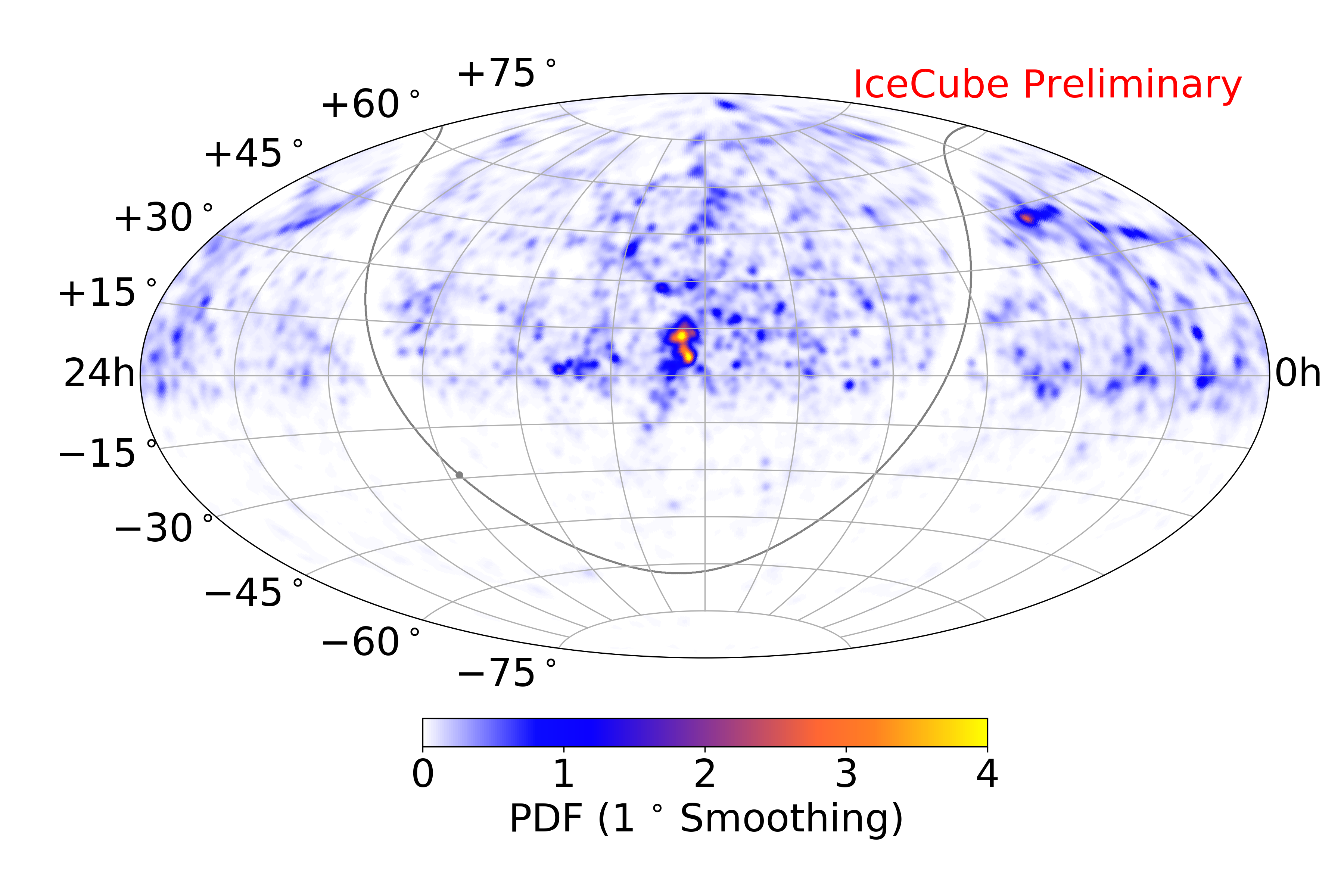} \\
a) Full 2MRS Catalog & b) Full 2MRS Catalog, weighted by redshift \\
\end{tabular}
\caption{Signal spatial PDF for a) Full 2MRS Catalog, and b) Full 2MRS Catalog with each galaxy weighted by its redshift.  Galaxies located in the center of the map make up the Virgo Supercluster and are assigned larger weight when redshift information is included.}
\label{fig:pdfs}
\end{figure} 

 Three different signal hypotheses are tested in this analysis. These correspond to:\\
 \\
$\bullet$ All galaxies in the 2MRS Catalog \\
$\bullet$ All galaxies in the 2MRS Catalog with $z$ \textless \,0.03 \\
$\bullet$ All galaxies in the 2MRS Catalog each galaxy weighted by its redshift distance  \\

A template PDF is created for each signal hypothesis, examples of which are shown in Figure \ref{fig:pdfs}. The template with a cutoff at $z$=0.03 corresponds to a close but relatively complete catalog. The redshift-weighted catalog includes weighting for each galaxy proportional to the inverse square of distance where distance is calculated from the redshift. This does not include any corrections for galaxy peculiar velocity effects or redshift uncertainty. The full catalog uses the complete survey.\\

Galaxies are binned using HEALPix bins of 0.2 square degrees \cite{healpix}. The PDF is based on the galaxy density map, convolved with the detector acceptance for an $E^{-2}$ hypothesis as a function of event declination. The resulting map is then convolved with a Gaussian of width equal to each event angular uncertainty.

\section{Results}
The p-values reported in Table \ref{table:results} denote the probability that the observed TS can be caused by a chance coincidence. All three scenarios are consistent with the no-correlation hypothesis.  Given the absence of a significant correlation, limits are placed on the neutrino flux that originates from interactions between diffuse intergalactic UHECR and matter in local galaxies.  These limits are reported for each hypothesis.  \\

\begin{table}[htb]
\begin{center}
\begin{tabular}{|c|c|c|c|} \hline 

Template & Test Statistic &p-value & \begin{tabular}[x]{@{}c@{}}Upper Limit $\Phi_{90\%}$ \\dN/dE (GeV$^{-1}$ s$^{-1}$cm$^{-2}$)\end{tabular} \\
\hline
Full Catalog Template  & 0.0 & 1.0 & 2.89 $\times$ 10$^{-18}$ \\
\hline
$z$ \textless 0.03 Template & 0.0 & 1.0 & 2.15 $\times$ 10$^{-18}$ \\
\hline
\begin{tabular}[x]{@{}c@{}}Full Catalog Template  \\with redshift weighting\end{tabular}& 0.0 & 1.0  & 1.97 $\times$ 10$^{-18}$ \\
\hline
\end{tabular} 
\end{center}
\caption{Analysis TS and p-value results.  Fluxes are integrated over the full sky and parameterized as dN/dE = $\Phi_{90\%}$ $\times$ ($\frac{E}{100 TeV}$)$^{-2}$ GeV$^{-1}$cm$^{-2}$s$^{-1}$ with 90$\%$ confidence level upper limits.}
\label{table:results}
\end{table}

\begin{figure}[htb]
\centering
\includegraphics[width=.75\textwidth]{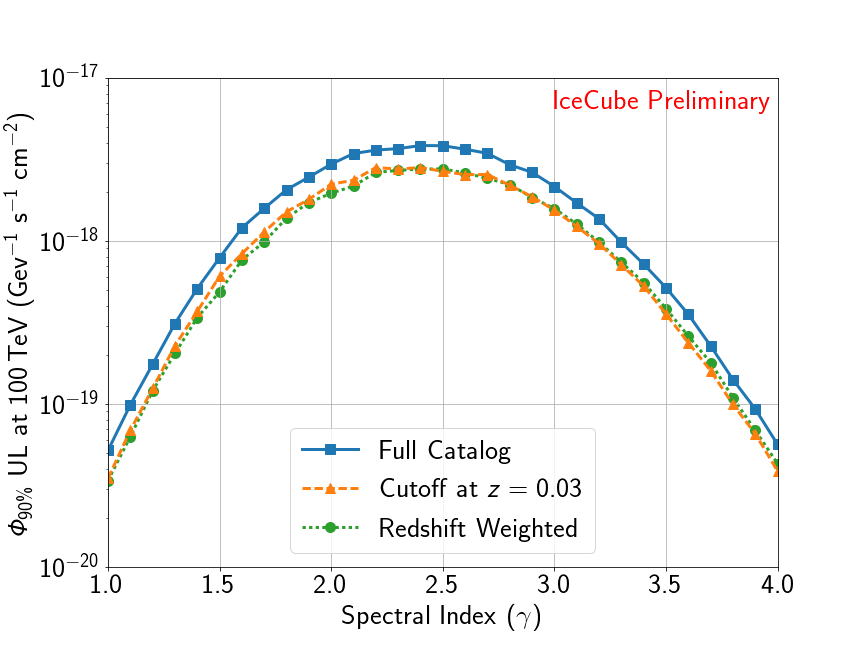} \\
\caption{Flux Upper limits for a range of source spectrum and templates.  Flux greater than these points can be excluded with a 90\% confidence level}
\label{fig:UL}
\end{figure}

These limits are calculated based on an assumed $E^{-2}$ energy spectrum. They are also computed for a source energy distribution that follow different spectra.  The 90\% confidence limits for other spectra are shown in Figure \ref{fig:UL} for each template.  \\

This analysis is limited by assumptions of the column density, distance, and completeness of the galaxies in the 2MRS survey.  As additional surveys like the LSST provide a more complete view of the local universe, this analysis could be expanded to include other available information like galaxy luminosity, and provide better limits on the relationship between neutrinos and local matter.

\end{document}